\PassOptionsToPackage{final}{graphicx}
\PassOptionsToPackage{final}{listings}
\documentclass[paper]{llncs}
\usepackage[title]{appendix}

\usepackage{hyperref}       
\usepackage{url}            
\usepackage{amsmath,amssymb,amsfonts}       
\usepackage{booktabs}       
\usepackage{xcolor}
\usepackage{xspace}
\usepackage{xfrac} 
\usepackage{paralist}
\usepackage[capitalize]{cleveref}
\usepackage{datetime2}
\usepackage{subcaption}
\usepackage{cases}
\usepackage{pgfplots}
\pgfplotsset{compat=newest}
\usepackage{wrapfig}
\usetikzlibrary{fit}

\usepackage{sansmath}
\usepackage{amssymb}
\usepackage[ruled,vlined]{algorithm2e}
\usepackage[inline]{enumitem}
\usepackage{amsmath}
\usepackage{amsfonts}
\usepackage{amsopn}
\usepackage{amscd}
\usepackage{hyperref}
\usepackage{verbatim}
\usepackage{tikz}
\usepackage{proof}
\usepackage{alltt}
\usepackage{color}
\usepackage{xspace}
\usepackage{tikz,ifthen}
\usepackage{fancybox}
\usepackage{wrapfig}
\usepackage{multirow}
\usepackage{listings}
\usepackage{rotating}
\usepackage{xcolor,colortbl}
\usepackage[textsize=small,colorinlistoftodos]{todonotes}
\usepackage[outdir=./images/]{epstopdf}
\usepackage{sansmath}
\usepackage{mathtools}

\usepackage{graphicx}
\usepackage{pxfonts}
\usepackage{caption}
\usepackage{subcaption}

\newcommand{\act}[1][a]{\alpha} 

\newcommand{\ltp}[1][L]{\mathcal{L}}   

\newcommand{\bddstate}[1]{s_{\mathbb{B}}{}}

\newcommand{\tool}[1]{\texttt{#1}\xspace}

\newcommand{\prism}{\tool{PRISM}}
\newcommand{\prismgames}{\tool{PRISM-games}}
\newcommand{\storm}{\tool{Storm}}

\newcommand{\tempest}{\tool{Tempest}}

\newcommand{\gist}{\tool{GIST}}
\newcommand{\uppaaltiga}{\tool{UPPAAL-Tiga}}
\newcommand{\quasy}{\tool{QUASY}}

\newcommand{\presafety}{\tool{PreSafety}}
\newcommand{\postsafety}{\tool{PostSafety}}
\newcommand{\optimal}{\tool{Optimal}}

\DeclareRobustCommand{\cpp}
{\valign{\vfil\hbox{##}\vfil\cr
   \textsf{C\kern-.1em}\cr
   $\hbox{\fontsize{\sf@size}{0}\textbf{+\kern-0.05em+}}$\cr}%
}

\selectcolormodel{cmy}

\definecolor{lgrey}{rgb}{0.8,0.8,0.8}
\definecolor{grey}{rgb}{0.5,0.5,0.5}

\definecolor{lightblue}{rgb}{0.8,0.8,1.0}
\definecolor{lightred}{rgb}{1.0,0.8,0.8}
\definecolor{lightgreen}{rgb}{0.8,1.0,0.8}
\definecolor{angrygreen}{cmyk}{0.279,0,0.91,0.08}
\definecolor{lightred}{rgb}{1.0,0.8,0.8}
\definecolor{pink}{rgb}{1.0,0.1,1.0}

\definecolor{prismgreen}{HTML}{009900}
\definecolor{prismred}{HTML}{cc0000}
\definecolor{prismblue}{HTML}{0000cc}

\usepackage{listings}
\usepackage{xcolor}

\definecolor{codegreen}{rgb}{0,0.6,0}
\definecolor{codegray}{rgb}{0.5,0.5,0.5}
\definecolor{codepurple}{rgb}{0.58,0,0.82}
\definecolor{backcolour}{rgb}{0.98,0.98,0.96}

\lstdefinestyle{mystyle}{
    backgroundcolor=\color{backcolour},
    commentstyle=\color{codegreen},
    keywordstyle=\color{magenta},
    numberstyle=\tiny\color{codegray},
    stringstyle=\color{codepurple},
    basicstyle=\ttfamily\footnotesize,
    breakatwhitespace=false,
    breaklines=true,
    captionpos=b,
    keepspaces=true,
    numbers=left,
    numbersep=5pt,
    showspaces=false,
    showstringspaces=false,
    showtabs=false,
    tabsize=2
}

\title{TEMPEST - Synthesis Tool for Reactive Systems and Shields in Probabilistic Environments\thanks{This project has received funding from the European Union’s Horizon 2020 research and innovation programme under grant agreement N° 956123 - FOCETA.}}
\author{
Stefan Pranger\inst{1} \and
Bettina Könighofer\inst{1,2}  \and
Lukas Posch\inst{1} \and
Roderick Bloem\inst{1,2}}

\institute{
Graz University of Technology, Institute IAIK, Austria
\and
Silicon Austria Labs, TU Graz SAL-DES Lab, Austria
}

\makeatletter
\def\thanks#1{\protected@xdef\@thanks{\@thanks
        \protect\footnotetext{#1}}}
\makeatother

\begin{document}

\maketitle

\begin{abstract}
	
We present \tempest, a synthesis tool to automatically create correct-by-construction reactive systems and shields
from qualitative or quantitative specifications in probabilistic environments.
A shield is a special type of reactive system used for run-time enforcement; i.e., a shield enforces a given qualitative or quantitative specification of a running system while interfering with its operation as little as possible.
 Shields that enforce a qualitative or quantitative specification are called safety-shields
or optimal-shields, respectively. Safety-shields can be implemented as pre-shields or as post-shields,
optimal-shields are implemented as post-shields.
Pre-shields are placed before the system and restrict the choices of the system. Post-shields are implemented after the system and are able to overwrite  the  system’s  output.
\tempest is based on the probabilistic model checker \storm, adding model checking algorithms for stochastic games
with safety and mean-payoff objectives.
To the best of our knowledge, \tempest is the only synthesis tool able to
solve 2$\sfrac{1}{2}$-player games with mean-payoff objectives without restrictions on the state space.
Furthermore, \tempest adds the functionality to synthesize safe and optimal strategies
that implement reactive systems and shields.
\end{abstract}

\setlength{\intextsep}{0pt}%
\section{Introduction}

\begin{figure}[tb]
\begin{minipage}[t]{0.48\textwidth}
\begin{tikzpicture}
\definecolor{myblack}{cmyk}{.67,.33,0,.99}
\tikzstyle{box} = [rectangle, rounded corners, minimum height=0.8cm, text centered, draw=myblack]
\tikzstyle{arrow} = [thick,->,>=stealth]
\tikzstyle{point} = [coordinate]
\definecolor{dgreen}{cmyk}{60, 0,100,0}

\node (env) [box] {Environment};
\node[inner sep=1pt, right of=env, xshift=2.75cm, yshift=1.62cm] (shield) {{{\includegraphics[bb= 15 10 800 780,scale=0.06]{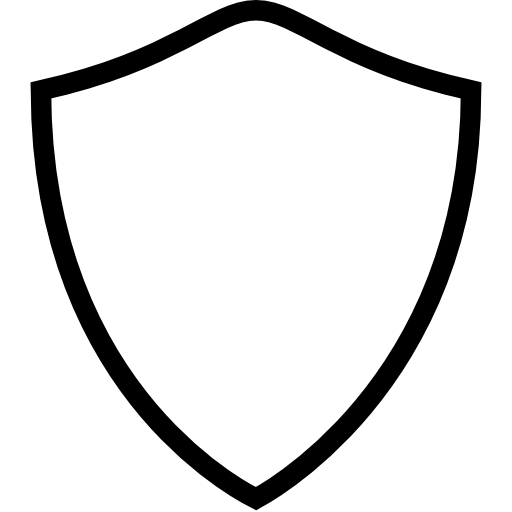}}}};
\node (system) [box,  right of=env, xshift=2.4cm] {System};

\draw [arrow, thick] (1.05, 0) -- node [anchor=south]{} (2.8,0);

\draw[thick] (1.2,0) --  (1.2,1.2);
\draw [arrow, thick] (1.2,1.2) |- node[anchor=north] {} (3,1.2);
\node[text width=1cm] at (2,1.5) {input};

\draw[arrow, thick] (3.42, 0.8) -- node [anchor=east]{$\{$actions$\}~~$} (3.42,0.4);

\draw[thick] (3.42, -0.4) --  (3.42,-0.7);

\draw [arrow, thick] (3.42,-0.7) -| node[pos=0.2, above] {action} (env);

\node[text width=1cm] at (3.7,1.3) {$Sh$};

\end{tikzpicture}
\end{minipage}
\begin{minipage}[t]{0.48\textwidth}
\begin{tikzpicture}
\definecolor{myblack}{cmyk}{.67,.33,0,.99}
\tikzstyle{box} = [rectangle, rounded corners, text width=2cm, minimum height=0.8cm, text centered, draw=myblack]
\tikzstyle{arrow} = [thick,->,>=stealth]
\tikzstyle{point} = [coordinate]
\definecolor{dgreen}{cmyk}{60, 0,100,0}

\node (env) [box] {Environment};
\node (system) [box, text width=1cm, right of=env, xshift=2.4cm] {System};


\node[inner sep=1pt, right of=env, xshift=2.73cm, yshift=-1.23cm] (shield) {{{\includegraphics[bb= 15 10 800 780,scale=0.06]{pics/shield-icon-blank}}}};

\draw [arrow, thick] (1.05, 0) -- node [anchor=south]{} (2.8,0);
\draw [arrow, thick] (3.42,-2.2) -| node[pos=0.3, above] {action'} (env);
\draw [arrow, thick] (3.42, -0.4) -- node [anchor=west]{action} (3.42, -0.95);
\draw[thick] (3.42, -2.05) --  (3.42,-2.2);
\node[text width=1cm] at (1.8,-0.5) {input};
\draw[thick] (1.2, 0) --  (1.2,-1.5);
\draw [arrow, thick] (1.2,-1.5) |- node[anchor=north] {} (2.9,-1.5);
\node[text width=1cm] at (3.7,-1.5) {$Sh$};

\end{tikzpicture}
\end{minipage}
    \caption{\emph{Left:} Pre-shielding. \emph{Right:} Post-shielding.}
    \label{fig:shielding_framework}
\end{figure}

\emph{Reactive synthesis} aims to automatically construct correct and efficient systems w.r.t. a formal specification
and has been increasingly used in a wide range of safety-critical applications.
%
A natural model for reactive synthesis is to model some inputs from the environment probabilistically
and some adverserially.
For adverserial inputs, the synthesized system assumes the worst case, for probabilistic inputs the average case.
The corresponding synthesis problem is mapped to solving a \emph{competitive stochastic turn-based game}, i.e., a
$2\sfrac{1}{2}$-player game. \emph{Qualitative specifications} specify the functional requirements of reactive systems.
With a \emph{quantitative specification} such as \emph{mean-payoff objectives}, we can measure how well a system satisfies the specification.
\emph{Shield synthesis} defines a synthesis framework to construct run-time enforcement modules called \emph{shields}
to guarantee the correctness of running systems. The concept of shielding is very general.
Shields that enforce qualitative objectives are so-called \emph{safety-shields}~\cite{AlshiekhBEKNT18}. For safety-shields, we
distinguish between pre-shielding and post-shielding as depicted in Fig.~\ref{fig:shielding_framework}.
In \emph{pre-shielding} the shield is implemented before the system
and restricts the choices for the system to a list of correct actions.
Pre-shielding is becoming increasingly important in the setting of safe reinforcement learning~\cite{0001KJSB20}.
In \emph{post-shielding}, the shield monitors the actions selected by the system and corrects them
if the chosen action could lead to a specification violation.
Shields that enforce quantitative measures are called \emph{optimal-shields}~\cite{DBLP:conf/cav/AvniBCHKP19}
and are implemented as post-shields. 
\tempest is able to synthesize optimal-shields that enforce a mean-payoff objective.
Optimal-shields that enforce multiple quantitative objectives can be obtained via a linear combination to give an approximate solution of a single mean-payoff objective. For instance, the decision whether an optimal-shield should interfere
could be based on first, a \emph{performance objective} to be minimized by the shield, and second, an \emph{interference cost} for changing the output of the system. An optimal-shield can then be computed \emph{by minimizing a single mean-payoff objective} obtained by combining both measures, thus guaranteeing maximal performance with minimal interference.

\textbf{Tempest capabilities.}
The core functionality of \tempest is the synthesis of reactive systems and shields in environments that incorporate uncertainty. To the best of our knowledge, \tempest is the only tool able to solve 2$\sfrac{1}{2}$-player games with mean-payoff objectives and qualitative objectives given in probabilistic temporal logics, without any restrictions on the state space.
Furthermore, \tempest is designed as synthesis tool. Therefore, the computed strategies
can intuitively be used as the synthesized system, which is not the case for many game-solving tools.
\tempest is the first tool available for the synthesis of shields
and is able to synthesize pre-safety and post-safety-shields, and optimal-shields.

\textbf{Implementation and availability.}
The tool is written in C++ and
builds upon the code-base of the model checker \storm~\cite{DJKV17},
extending existing features to provide the capability of solving stochastic games.
\tempest is available under the GPL-3 open source license.
The tool and its source code, along with a docker image and several examples,
are available from the \tempest web page\footnote{\url{http://www.tempest-shielding.xyz}}.

\textbf{Connections to other tools.}
Probabilistic model checking tools like \prism~\cite{Prism4} and \storm~\cite{DJKV17} provide verification of quantitative reward-based  properties and qualitative properties in probabilistic temporal logics.
Many synthesis tools based on games are available and widely used, for example, \uppaaltiga~\cite{BehrmannCDFLL07} is able to solve qualitative timed games, \gist~\cite{10.1007/978-3-642-14295-6_57} solves qualitative stochastic games,
and \quasy~\cite{QUASY} solves mean-payoff 2-player games.
\prismgames~3.0~\cite{Prism-Games} is able to solve turn-based stochastic multi-player games
under a variety of properties including long-run average~\cite{ChenFKPS13}.
However, for solving long-run average objectives, \prismgames needs the game to be a \textit{controllable multi-chain}, i.e., one of the players needs to be able to reach every end-component from any state with probability one. \textit{This is a strong assumption on the structure of the game graph, which many models used in synthesis do not fulfill.}
In contrast, \tempest does not rely on any assumptions on the structure of the game graph. 




\textbf{Acknowledgements.} We would like to thank both Tim Quatmann and Joost-Pieter Katoen for their continuous help on getting acquainted with the source code of \storm.

\section{Model and Property Specification}

\textbf{\tempest Model Specification.}
\tempest supports turn-based \emph{stochastic multi-player games} (SMGs)
and uses PRISM-games's modelling language to describe the game~\cite{Prism-Games}.
The players are divided in two competing coalitions, where the first team is working together to satisfy or maximize/minimize a property
given in rPATL~\cite{ChenFKPS13}.
In each state,  exactly one player chooses an available probabilistic transition to determine the next state.
A \emph{strategy} for a player determines the choices of transitions made by the player.

\noindent\textbf{\tempest Property Specification.}
For \emph{the synthesis of reactive systems}, \tempest uses the property specification language of PRISM-games to express properties in
rPATL~\cite{ChenFKPS13}.
We give a few examples that can be used in \tempest:
\begin{itemize}
    \item $\langle\langle 1, 2\rangle\rangle P_{max=?} [F~target]$:
    Using the operator $P_{max=?}$, \tempest computes a strategy for the  player coalition of player 1 and 2 that guarantees
    to reach \emph{target} with the largest probability.
    \item $\langle\langle 1, 2\rangle\rangle R^r_{max=?} [ S ]$: Using the operators $R$ and $S$, \tempest synthesizes
    a strategy that maximizes the expected averaged reward $r$ in the \emph{long-run}.
\end{itemize}

\textbf{Synthesis of safety-shields.}
For the synthesis of shields, \tempest extends the property specification language.
Let the \emph{safety-value} of an action in a certain state be the maximal
probability to stay safe within the next $k$ steps when executing this action.
A safety-shield decides whether an action is blocked in a certain state based
on either an \emph{absolute} threshold $\gamma$, or a relative threshold $\lambda$.
A shield using an absolute threshold blocks all actions with a safety-value smaller than $\gamma$.
By using a relative threshold, actions are blocked with a safety-value smaller than
the best safety-value of an action in the current state times $\lambda$.
The syntax for safety-shielding requires to
specify the type of shielding using the keywords \presafety and \postsafety,
and to define the used threshold.
We give the following examples:
\begin{itemize}
    \item $ \langle \text{PreSafety}, \gamma=0.9 \rangle \langle \langle shield \rangle \rangle P_{max=?} [ G^{<=14} ~! crash ] $:
    Using this property, \tempest synthesizes a pre-safety-shield that allows all actions that do not cause a crash with
    a maximal probability of $0.9$ within the next $14$ time steps.
    \item $ \langle \text{PostSafety}, \lambda=0.95 \rangle \langle \langle shield \rangle \rangle P_{max=?} [ G^{<=14} ~! crash ] $:
    \tempest synthesizes a post-safety-shield that blocks all actions using the relative threshold $\lambda$.
\end{itemize}

\textbf{Synthesis of optimal-shields.} The property 
starts with the keyword \optimal followed by the expression used to compute the long run average. For example:
\begin{itemize}
    \item $ \langle \text{Optimal} \rangle \langle \langle shield \rangle \rangle R^r_{min=?} [ S ] $: \tempest computes an optimal shield that guarantees the long-run average reward of $r$.
\end{itemize}

\section{\tempest Synthesis of Strategies}

\tempest computes a memoryless deterministic strategy, implementing
a reactive system or a shield, under which
the specified property can be guaranteed. The strategy is computed using \emph{value iteration}
to solve the coalition game. Fig.~\ref{fig:shield_output} shows sample outputs of the strategies
of the first experiment given in Section~\ref{sec:warehouse}, implementing pre-safety and post-safety-shields.
In the pre-shielding case, the strategy provides for any state a list of allowed actions
with its corresponding safety-value.
The strategy for post-shielding defines for every state and available action, the action to be 
forwarded by the shield.



\lstset{
  escapechar={|}
}
\begin{figure}
\centering
\begin{lstlisting}[mathescape,language=Python, label= listing_pre_post, basicstyle=\fontsize{7.5}{8}\selectfont\ttfamily]
|\color{green}Pre-Safety-Shield with absolute comparison (gamma = 0.8):|
 |\color{green}state\_id [label]:  'allowed actions' [<value>: (<action\_id {label})>]:|

0 [move=0 & x1=0 & y1=0 & x2=4 & y2=4]:  1.0:(0 {e}); 1:(1 {s})
3 [move=0 & x1=1 & y1=0 & x2=3 & y2=4]:  0.9:(0 {e}); 1:(2 {w})
4 [move=0 & x1=1 & y1=0 & x2=4 & y2=4]:  0.9:(1 {s}); 1:(3 {n})
    .....
\end{lstlisting}
\end{figure}
\vspace{-2.0mm}
\begin{figure}
\centering
\begin{lstlisting}[mathescape,language=Python, label= listing_post, basicstyle=\fontsize{7.5}{8}\selectfont\ttfamily]
|\color{green}Post-Safety-Shield with relative comparison (lambda = 0.95):|
 |\color{green}state\_id [label]: 'forwarded actions' [<action\_id> {label}: <forwarded\_action\_id> {label}]:|

0 [move=0 & x1=0 & y1=0 & x2=4	& y2=4]:  0{e}:0{e}; 1{s}:1{s}
3 [move=0 & x1=1 & y1=0 & x2=3	& y2=4]:  0{e}:2{w}; 2{w}:2{w}
4 [move=0 & x1=1 & y1=0 & x2=4	& y2=4]:  1{s}:3{n}; 3{n}:3{n}
    .....

\end{lstlisting}
    \caption{Synthesized strategies implementing a pre-safety-shield (\emph{top}) and a post-safety-shield (\emph{bottom}).}
    \label{fig:shield_output}
\end{figure}

\section{\tempest in Action}





\label{sec:warehouse}

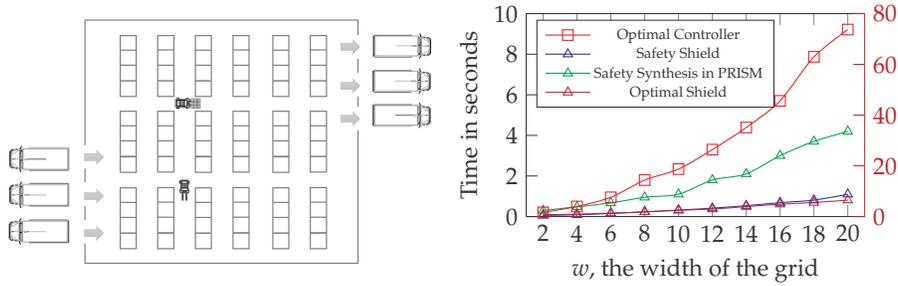
\begin{figure}[tb]
\begin{minipage}[t]{0.48\textwidth}

\includegraphics[width=5.7cm]{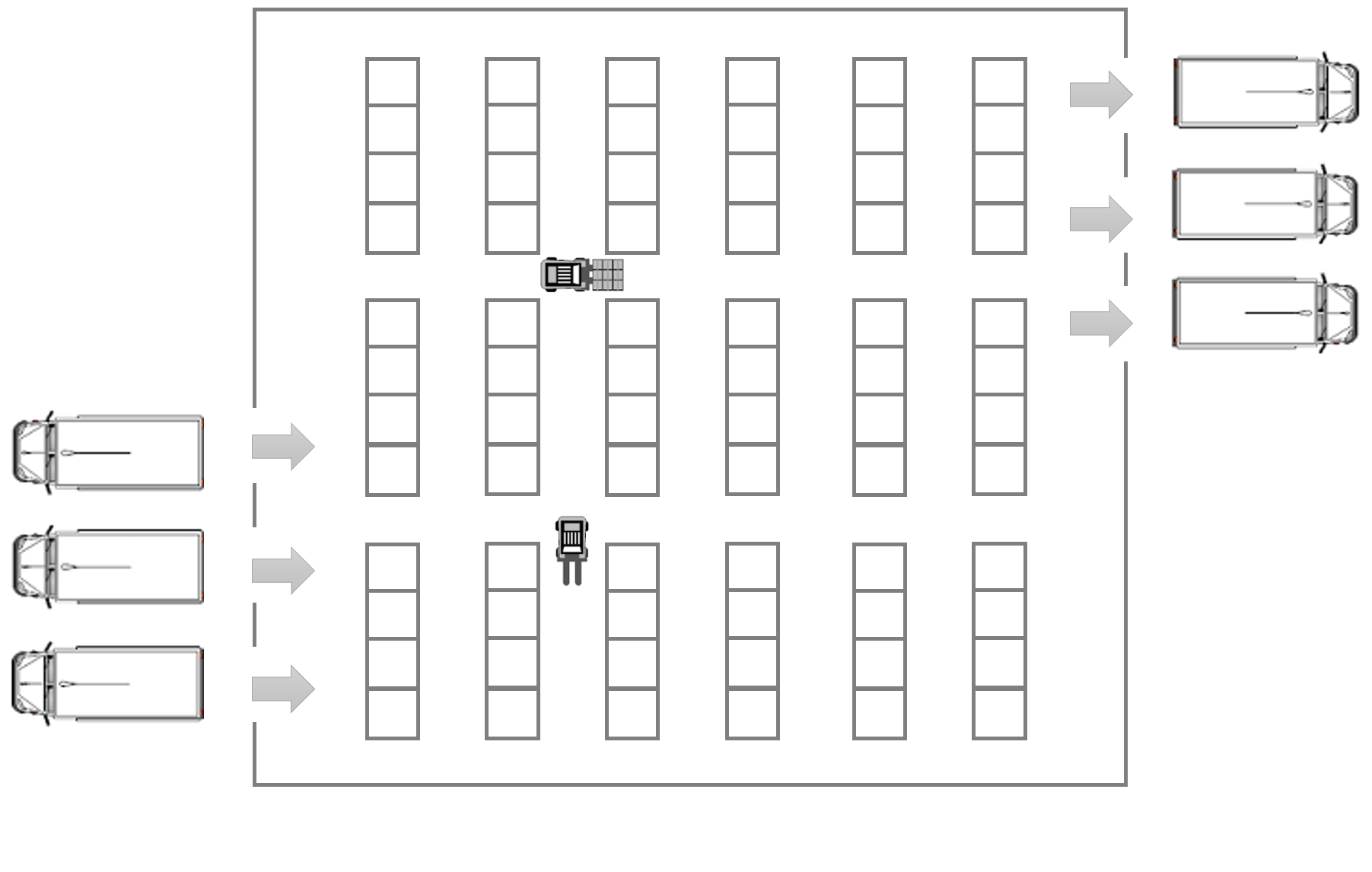}
\end{minipage}
\begin{minipage}[t]{0.48\textwidth}
  \begin{tikzpicture}
    \pgfplotsset{
      scale only axis,
      xtick={2,4,6,8,10,12,14,16,18,20},
      xmin=1, xmax=21,
      y axis style/.style={
        yticklabel style=#1,
        ylabel style=#1,
        y axis line style=#1,
        ytick style=#1
      }
    }
    \begin{axis}[
      axis y line*=left,
      y axis style=black!75!black,
      ymin=0, ymax=10,
      xlabel={$w$, the width of the grid},
      ylabel=Time in seconds,
      width=4.5cm,
      height=2.7cm
    ]
    \addplot[smooth,mark=triangle,blue]
      coordinates{
        (2, 0.058)
        (4, 0.089)
        (6, 0.153)
        (8, 0.232)
        (10, 0.321)
        (12, 0.414)
        (14, 0.535)
        (16, 0.686)
        (18, 0.809)
        (20, 1.1)
    };\label{Safety-Shield}
    \addplot[smooth,mark=triangle,green]
      coordinates{
        (2, 0.3)
        (4, 0.48)
        (6, 0.67)
        (8, 0.95)
        (10, 1.1)
        (12, 1.8)
        (14, 2.1)
        (16, 3)
        (18, 3.7)
        (20, 4.2)
    };\label{Safety-Synthesis-in-PRISM}
    \addplot[smooth,mark=triangle,purple]
      coordinates{
        (2, 0.09)
        (4, 0.12)
        (6, 0.17)
        (8, 0.232)
        (10, 0.31)
        (12, 0.37)
        (14, 0.49)
        (16, 0.61)
        (18, 0.69)
        (20, 0.81)
    };\label{Optimal-Shield}
    \end{axis}
    \begin{axis}[
      axis y line*=right,
      axis x line=none,
      ymin=0, ymax=80,
      y axis style=red!75!black,
      width=4.5cm,
      height=2.7cm,
      legend pos=north west,
      legend style={nodes={scale=0.61, transform shape}},
    ]
    \addplot[smooth,mark=square,red]
      coordinates{
          (2, 1.645)
          (4, 3.983)
          (6, 7.584)
          (8, 14.38)
          (10, 18.7)
          (12, 26.39)
          (14, 35.11)
          (16, 45.61)
          (18, 62.96)
          (20, 73.71)
    };\addlegendentry{Optimal Controller}
    \addlegendimage{/pgfplots/refstyle=Safety-Shield}\addlegendentry{Safety Shield}
    \addlegendimage{/pgfplots/refstyle=Safety-Synthesis-in-PRISM}\addlegendentry{Safety Synthesis in PRISM}
    \addlegendimage{/pgfplots/refstyle=Optimal-Shield}\addlegendentry{Optimal Shield}
    \end{axis}
  \end{tikzpicture}
\end{minipage}
   \vspace{-0.5cm}
    \caption{\emph{Left:} Warehouse floor plan with $6~\times~3$ shelves. \emph{Right:} Synthesis-times for
    controller synthesis, safety-shield synthesis and optimal-shield synthesis.}
    \label{fig:warehouse}
\end{figure}

\textbf{High-Level Planning in Robotics.}
A classical application of reactive synthesis is the domain of automated high-level 
planning in robotics. We consider a warehouse floor plan with several
shelves, see Fig.~\ref{fig:warehouse} (left). To parametrise the experiment, we consider
floor plans with  $n~\times~3$ shelves with $2\leq n \leq 20$. 
A robot operates together among other robots within the warehouse.
\tempest can be used in this setting to synthesize controllers for the robot that perform certain tasks,
as well as shields used to ensure safe operation of the robot, or to guarantee performance.
\textbf{Controller synthesis:}
Using \tempest, we synthesize a controller for the robot that repeatedly picks up packages from one of the entrances and 
delivers them to the exits. 
We use the mean-payoff criterion to specify that the stochastic shortest paths should be taken.
%
\textbf{Safety-shield synthesis:}
A safety-shield can be synthesized to enforce collision avoidance with other robots.
In the experiments, we used a finite horizon of $14$ steps and a relative threshold of $\lambda=0.9$.
\textbf{Optimal-shield synthesis:}
During operation, a corridor may be blocked.
A robot should not unnecessarily wait for the corridor to be traversable when alternative paths exist.
We synthesize an optimal-shield that penalizes 'waiting' and is able to enforce a detour when waiting
gets too expensive.
\textbf{Results.} The models, parameters, and properties used for all experiments can be found on the \tempest website.
The results for the synthesis-times are depicted in Fig.~\ref{fig:warehouse} (right).
The sizes of state space of the game graphs range from $5184$ states for $n=2$ to $186624$ states for $n=20$.
The results for optimal-shields use the axis on the right hand-side.
The times for creating pre-safety and post-safety-shields are identical.
To compare our results, we tried to compute a strategy for the optimal controller in \prismgames~3.0
which resulted in an error. By proper modelling, we were able to synthesize safe controllers comparable to the
safety-shield using \prismgames, resulting in better synthesis-times for \tempest.

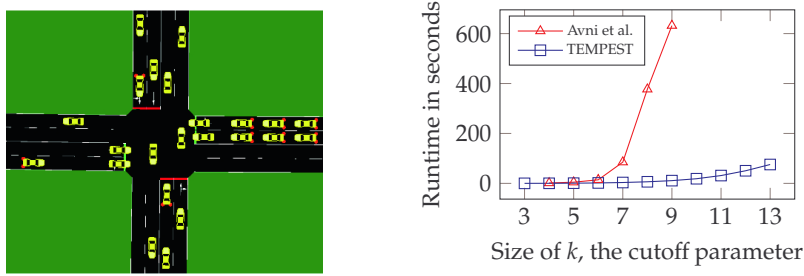
\begin{figure}[tb]
\begin{minipage}[t]{0.49\textwidth}
\centering
\includegraphics[width=4.2cm]{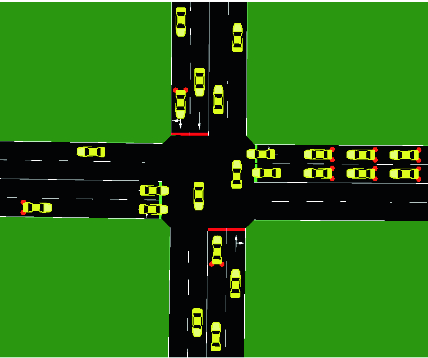}
\end{minipage}
\begin{minipage}[t]{0.48\textwidth}
\centering
  \begin{tikzpicture}
    \begin{axis}[
      xlabel={Size of $k$, the cutoff parameter},
      ylabel={Runtime in seconds},
      xtick={3,5,7,9,11,13},
      legend style={nodes={scale=0.62, transform shape}},
      legend pos=north west,
      grid style=dashed,
      width=5.5cm,
      height=4.1cm
  ]
      \addplot [
        color=red,
        mark=triangle,
          ]coordinates {
        (4, 1.5)
        (5, 4.7)
        (6, 14.3)
        (7, 85)
        (8, 377)
        (9, 632)
      };
      \addlegendentry{Avni et al.}
      \addplot [
        color=blue,
        mark=square,
          ]coordinates {
            (3, 0.11)
            (4, 0.319)
            (5, 0.825)
            (6, 1.71)
            (7, 3.44)
            (8, 6.35)
            (9, 10.955)
            (10, 18.59)
            (11, 31.3)
            (12, 50.494)
            (13, 75.587)
      };
      \addlegendentry{TEMPEST}
    \end{axis}
  \end{tikzpicture}
\end{minipage}
    \caption{\emph{Left:} Synthesis times for safety-shield.
    \emph{Right:} Comparison of synthesis times for optimal-shields: \tempest vs Avni et.al.'s implementation~\cite{DBLP:conf/cav/AvniBCHKP19}.}
    \label{fig:experiment_2_3}
\end{figure}

\noindent\textbf{Optimal-Shielding  in Urban  Traffic  Control.}
Avni et al.~\cite{DBLP:conf/cav/AvniBCHKP19} synthesize optimal-shields 
that overwrite the commands of a traffic-light controller modeled in the traffic simulator
SUMO. 
The optimal-shield needs to balance the number of waiting cars per incoming road
with the cost for interfering with the traffic light controller. 
The example is parametrised with the cut-off parameter $k$ that defines the 
maximal modelled number of waiting cars per road. 
The comparison of the synthesis-times from Avni et.al.’s  implementation and \tempest are shown in 
Fig.~\ref{fig:experiment_2_3} (right), showing a difference by orders of magnitudes
in favor of \tempest.


\section{Conclusion and Future Work}\label{sec:conclusion}

We have introduced \tempest, a tool for the synthesis of reactive systems and shields 
with properties given in rPATL capturing qualitative and quantitative objectives
with probabilities. 
Currently, \tempest supports perfect-information SMGs with full information.
In future work, we will investigate in efficient techniques to deal with partial information. 
Furthermore, we will extend \tempest's synthesis algorithm to support strategies with finite memory
with deterministic and stochastic-updates.


\bibliographystyle{abbrv}
\bibliography{literature}

\end{document}